\documentclass[prl,twocolumn,showpacs]{revtex4}
\usepackage{graphicx}
\usepackage{graphics}
\usepackage{amsfonts}
\usepackage{amsmath}
\usepackage{epsfig}

\usepackage{bm}

\psfigdriver{dvips}

\begin{document}
% define commands for international characters
\catcode`\ä = \active \catcode`\ö = \active \catcode`\ü = \active
\catcode`\Ä = \active \catcode`\Ö = \active \catcode`\Ü = \active
\catcode`\ß = \active \catcode`\é = \active \catcode`\è = \active
\catcode`\ë = \active \catcode`\ô = \active \catcode`\ê = \active
\catcode`\ø = \active \catcode`\ò = \active \catcode`\í = \active
\defä{\"a} \defö{\"o} \defü{\"u} \defÄ{\"A} \defÖ{\"O} \defÜ{\"U} \defß{\ss} \defé{\'{e}}
\defè{\`{e}} \defë{\"{e}} \defô{\^{o}} \defê{\^{e}} \defø{\o} \defò{\`{o}} \defí{\'{i}}
%\draft               % preprint mode
\newcommand{\li}{$^6$Li}
\newcommand{\na}{$^{23}$Na}
\newcommand{\vect}[1]{\mathbf #1}
\newcommand{\g}{g^{(2)}}
\newcommand{\one}{|1\rangle}
\newcommand{\two}{|2\rangle}
\newcommand{\V}{V_{12}}
% 2 col mode:
%@twocolumnfalse\endcsname %\vspace{-5mm}
\title{Spectroscopic insensitivity to cold collisions in a two-state mixture of fermions}

%\vspace{-5mm}

\author{Martin W. Zwierlein, Zoran Hadzibabic, Subhadeep Gupta, and Wolfgang Ketterle}

\affiliation{Department of Physics\mbox{,} MIT-Harvard Center for Ultracold Atoms\mbox{,} and Research Laboratory of Electronics,\\
MIT, Cambridge, MA 02139}

\date{June 24, 2003}

\begin{abstract}
We have experimentally demonstrated the absence of spectroscopic resonance shifts in a mixture of two interacting Fermi gases. This result is linked to observations in an ultracold gas of thermal bosons. There, the measured resonance shift due to interstate collisions is independent of the coherence in the system, and twice that expected from the equilibrium energy splitting between the two
%%% inserted: internal
internal states in a fully decohered cloud.
We give a simple theoretical explanation of these observations, which elucidates the effect of coherent radiation on an incoherent mixture of atoms.
\end{abstract}

\maketitle

%\narrowtext

%--- general introduction ------------------------------------

The coherence properties of light and matter are intimately
connected with the quantum statistics of the constituent
particles. One quantitative measure of the coherence in a system
is the two-particle correlation function at zero distance, $\g$,
which measures the
%conditional
probability that two
particles are simultaneously detected.
Intensity
fluctuations in the incoherent light emitted by a light bulb lead
to photon ``bunching", making this probability twice higher than
in the coherent light of a laser. Identical fermions on the other
hand exhibit ``anti-bunching", making such a probability zero.

Interactions in ultracold atomic gases crucially depend on the
value of $\g$~\cite{kett97}. The reason is that s-wave scattering relies on particles
overlapping in real space. The interaction energy in a many-body
system is determined by coherent collisions, for which
the outgoing and the incoming two-particle states are identical.
Under this constraint, the two colliding particles can at most do
two things - either preserve their momenta, or exchange them. We
can thus distinguish four cases: (1) Two identical bosons in a
thermal gas can collide in both ways, corresponding to $\g=2$. (2) Two atoms in a Bose-Einstein
condensate (BEC) have the same momenta and cannot undergo the
exchange interaction. Here $\g=1$.
(3) Two distinguishable particles, fermions or bosons, also cannot
exchange their momenta because that would make the outgoing state
different from the incoming one. Again, $\g=1$. (4) Two identical fermions cannot collide at all, so $\g=0$. In all cases, the
mean-field energy of a particle with mass $m$ is given by $\g (4\pi \hbar^2 / m)
a n$, where $a$ is the s-wave scattering length, and $n$ is the density
of atoms it interacts with.

%%% Inserted: Mean field energies and therefore g(2) can be measured spectroscopically.
Mean field energies and therefore $\g$ can be measured spectroscopically.
In experiments on ultracold hydrogen, mean-field shifts of the
1S-2S two-photon transition were used to prove the existence of a
BEC~\cite{frie98}. However, quantitative interpretation of the
shifts led to a vivid theoretical discussion about the coherence
related ``factors of 2"~\cite{cote99,okte99,okte02,peth01}.
%%% changed: Eric Cornell's group at JILA
More recently, Harber {\it et~al.} performed Ramsey
spectroscopy in a two-component, thermal gas of $^{87}$Rb bosons
to measure $\g$ in the interstate collisional shift~\cite{harb02}.
Their measurements yielded $\g=2$, independent of the degree of
coherence between the two states. The spectroscopic results thus
seemed to correspond to the case of all particles being in an identical
coherent superposition of the two internal states, even though the
binary mixture was partially decohered and should have had a mean-field
energy corresponding to $1<\g<2$. The authors commented on this mystery~\cite{harb02ICAP}: ``it is a pleasure to note that a two-level system can still yield surprises, 75 years after
the advent of quantum mechanics." The mystery can be formally resolved using a quantum Boltzmann equation~\cite{okte02spin, will02, brad02, fuch02, fuch03}.

Here, we experimentally address the relation between coherence and
spectroscopic measurements in a binary mixture of ultracold {\it
fermions}. We demonstrate that shifts of spectroscopic lines
are absent even in a fully decohered binary mixture, in which the
particles are distinguishable, and the many-body
mean-field energy in the system has developed. We theoretically
show that this is a direct consequence of the coherent nature of
the RF {\it excitation}, which, in general, leads to a final state with $\g$ different from the initial state.

%%% inserted: which, in general, leads to a final state with g(2) different from the initial state.
%%% deleted: and is not dependent on the coherence of
%%% the {\it sample} on which spectroscopy is performed.

Our calculation intuitively explains
both our results for fermions, and the results for bosons of ref.~\cite{harb02}.
%%%changed: the JILA results for bosons

In a recent paper~\cite{gupt03}, we demonstrated the absence of
mean-field ``clock-shifts" in a coherent two-state superposition
of {\li} fermions. In this case, RF spectroscopy
was performed on a gas prepared purely in one internal state. Since an
RF pulse acts as a rotation in the two-state Hilbert space, all
the atoms stayed in an identical (superposition) state and could
not interact. As long as the fermionic atoms were
indistinguishable, $\g=0$, and the resonance was thus found to be
unperturbed at $\nu_0 = \frac{E_{12}}{h}$, where $E_{12}$ is the
energy difference between the internal states $|1\rangle$ and
$|2\rangle$.

However, once decoherence sets in, for example due to
inhomogeneous magnetic fields across the cloud, the spatial
overlap between atoms in different states grows and mean-field
energy density builds up:
\begin{eqnarray}
{\cal E}_{\rm int}({\bf r}) = \g V_{12}\, n_1({\bf r})\, n_2({\bf r}),\quad V_{12} =
\frac{4\pi\hbar^2}{m}a_{12},\label{eq:meanfield}
\end{eqnarray}
where $n_1$ and $n_2$ are the local densities of particles in states
$\one$ and $\two$, and $a_{12}$ is the interstate s-wave
scattering length.
%%% Inserted:
%%% Here decoherence means that off-diagonal matrix elements of the density matrix have vanished locally.  As a
%%% result, everywhere in the sample, atoms are no longer in ONE pure state, but occupy TWO ORTHOGONAL  states.
%%% Therefore, s-wave collisions are no longer suppressed by the Pauli principle.
Here decoherence means that off-diagonal matrix elements of the density matrix have vanished locally.  As a result, everywhere in the sample, atoms are no longer in one pure state, but occupy two orthogonal states, and s-wave collisions are no longer suppressed by the Pauli principle.
In a fully decohered cloud, we have a binary
mixture of two distinct species of atoms, with a mean-field energy
density ${\cal E}_{\rm int} = V_{12} n_1 n_2$. This interaction
changes the equilibrium energy level of atoms in state $\one$
($\two$) according to $\delta \mu_{1,2} = V_{12} n_{2,1}$. The
difference in equilibrium mean-field energy of the two states is then
\begin{eqnarray}
\Delta E_{\rm int} = \delta\mu_2 - \delta\mu_1 =  \V (n_1 - n_2).
\label{eq:energydiff}
\end{eqnarray}
This suggests~\cite{gupt03, harb02, harb02ICAP} that in a decohering
sample, the resonant frequency for population transfer between the
two states gradually changes from $\nu_{12} = \nu_0$ to $\nu_{12}
= \nu_0 + \frac{1}{h}\Delta E_{\rm int}$. Here, we show both experimentally
and theoretically that this conclusion is wrong, and that the
spectroscopic resonance frequency $\nu_{12}$ is
always the unperturbed frequency $\nu_0$.

Our experimental setup was described in~\cite{gupt03, hadz03}.
About $10^7$ fermionic {\li} atoms were confined in an optical dipole trap at a temperature of $35\, {\rm \mu K}$.
The two-level system under consideration is
formed by the two lowest ground state hyperfine levels,
$|1\rangle$ and $|2\rangle$, corresponding to
$|F,m_F\rangle = |1/2,1/2\rangle$ and $|1/2,-1/2\rangle$ in the
low field basis, respectively. A DC magnetic field of $B = 320 \rm G$ was
applied to the sample in order to tune the interstate scattering
length $a_{12}$ to a large value of $\sim -300 a_0$, where $a_0$ is
the Bohr radius~\cite{gupt03}. 

%%% deleted: At this field, electronic and
%%% nuclear spins are mostly decoupled, and the two states experience
%%% approximately the same Zeeman shift. This greatly reduces the
%%% effect of magnetic field fluctuations, although the differential
%%% shift is still linear in $B$, unlike in~\cite{harb02}.

We created a superposition of atoms in states $\one$ and $\two$
using a non-adiabatic RF sweep around the energy splitting of $74
\rm MHz$. As the sample decohered, efficient evaporative cooling set in,
confirming a large elastic scattering length. After 1 second, we
were left with a fully decohered mixture at a mean density $n =
5 \times 10^{13} {\rm cm}^{-3}$. The rate of the RF sweep
was adjusted so that after decoherence and cooling, $80\%$ of the atoms
were in state $\two$. The mean-field interaction should thus have
increased the energy splitting of the two levels by $h \delta\nu = \delta\mu_2 -
\delta\mu_1 = \V (n_1 - n_2) \approx h \times 10 \,{\rm kHz}$. Our
experiments involving a third state~\cite{gupt03} have confirmed the presence of such energy shifts,
%%% inserted: and prove that full decoherence has been reached.
and prove that full decoherence has been reached.

\begin{figure}[htbp]
  \includegraphics[width=0.8\columnwidth]{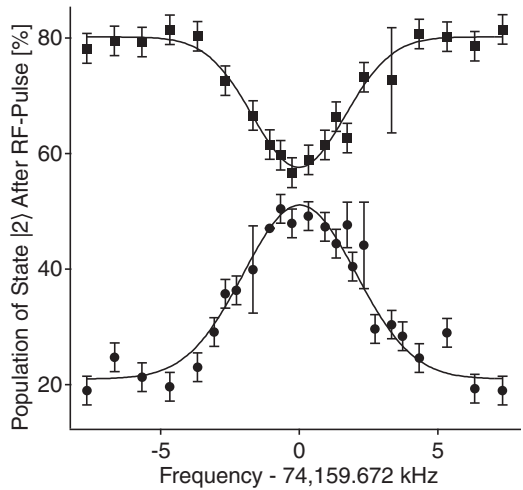}
  \caption{Absence of mean-field shift of an RF transition in a binary Fermi system. The resonance curves were measured for fully decohered 80\%/20\% two-state mixtures of fermions.
The measured frequency difference between the two lines is $(34
\pm 146)\,$Hz, even though Eq.~\ref{eq:energydiff} would predict a splitting of 20 kHz.}
\label{fig:result}
\end{figure}

Rabi spectroscopy in the interacting binary mixture was performed
by applying $200 \mu {\rm s}$ RF pulses of different frequencies,
and recording the final populations in the two states by
simultaneous absorption imaging (Fig.~\ref{fig:result}). In order
to eliminate the systematic uncertainty in the value of $\nu_0$,
we performed a second experiment with the population ratios of states $\one$
and $\two$ reversed. According to Eq.~\ref{eq:energydiff}, one would expect an opposite shift of the resonance.

%%% Changed: we performed a second experiment with the roles of states $\one$
%%% and $\two$ reversed. 

Within our precision, no interaction shift of the
resonance frequency was observed. Comparing the expected
difference in mean-field shifts for the two experiments, $2 \delta
\nu = 20\, {\rm kHz}$, with the measured line separation of $(34\,\pm\, 146)\,\rm Hz$, we arrive at an apparent value for $\g = 0.002(7)$. This demonstrates the universal absence of a
resonance shift in a very cold two-level Fermi gas, independent of the coherence in the system.

%%% This demonstrates the universal absence of a
%%% resonance shift in a two-level Fermi gas, independent of the coherence in the system.

Evidently, RF spectroscopy does not measure the expected
difference in thermodynamic chemical potentials for the two
states. Experiments with thermal bosons have posed a similar
puzzle~\cite{harb02}. Here we explain that this is a direct
consequence of the coherent nature of the RF excitation.

\begin{figure}[htbp]
  \includegraphics[width=\columnwidth]{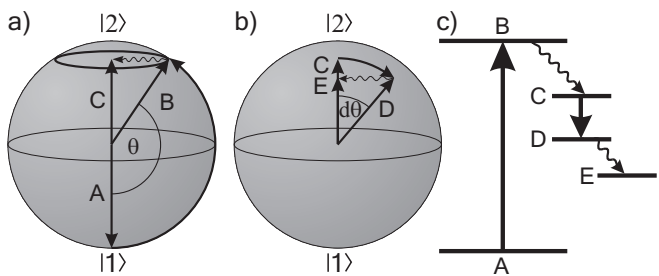}
  \caption{Bloch sphere representation of RF transitions. a) An RF pulse
rotates a pure state A into B. The superposition
state decoheres into a ``ring" distribution, represented by its average, C. b) A second RF pulse transforms the fully decohered state C into a partially coherent state D. The final state E is reached only after further decoherence. c) Transfers A$\rightarrow$B and C$\rightarrow$D are coherent and reversible. 
B$\rightarrow$C and D$\rightarrow$E are
irreversible.}
  \label{fig:bloch}
\end{figure}

In Fig.~\ref{fig:bloch}, the average properties of the many-body
state at a specific point ${\bf r}$ in the trap are described by the three coordinates of the local spin-1/2 Bloch vector $\vect{m}({\bf r}) = m_z({\bf r})  \hat{\vect{e}}_z + \vect{m}_\perp({\bf r})$. In the following, we omit the label ${\bf r}$. %%% included: at a specific point ${\bf r}$ in the trap
$m_z = \frac{n_2 - n_1}{2}$ represents the population difference in the two states, whereas the transverse component $\vect{m}_\perp$ is a measure of the coherence in the system. 
%%%included: In the following, we omit the label ${\bf r}$.
%In polar coordinates, 
The length of the Bloch vector measures the purity of the mixture %%% changed: mixture instead of state 
and hence the entropy of the system.
%The polar angle $\theta$ encodes the coherence of the sample.
Fully decohered statistical mixtures 
%%%inserted: do not have off-diagonal matrix elements of the density matrix and 
do not have off-diagonal matrix elements of the density matrix and
are represented by vectors with $\vect{m}_\perp = 0$,
with state A being the special case of a pure state. In Fig.~\ref{fig:bloch}a, state B
is created by applying an RF pulse on a pure sample A. In this case, there is no interaction energy in the system during the RF pulse, and no frequency shift is expected~\cite{gupt03}. State C
is formed through subsequent decoherence of state B. States B and C have the same number of particles in $\one$ and $\two$, but in C the mean-field has fully developed. 

Our experiment is performed on a C-like state (Fig.~\ref{fig:bloch}b). 
Here we explain why Eq.~\ref{eq:energydiff} still does not give the correct resonance frequency for an infinitesimal transfer of atoms between $\one$ and $\two$.
The key point is that even though the sample is fully decohered, the applied RF pulse re-introduces coherence into the system. 
%%% Inserted: According to Eq. (6) below, this can change the value of g(2).
According to Eq. (6) below, this will change the value of $\g$.
Let us consider two fully decohered states, C and E. Eq.~\ref{eq:energydiff} correctly gives the energy of the transformation C$\rightarrow$E. However, these two states have different entropies, as indicated by Bloch vectors of different lengths. An RF pulse is a unitary transformation of the system, and must preserve entropy.
The true effect of the RF pulse is thus to  change the relative
populations of $\one$ and $\two$ by tilting the Bloch vector away
from the $z$ axis, into state D. It is the energy of {\it this} transformation, C$\rightarrow$D, that
needs to be calculated in order to find the correct resonant RF frequency. 

%%% changed: In the case of fermions, we can prove

In the case of fermions
%%% inserted: with short-range (delta function) interactions
with short-range (delta function) interactions, we can prove
very generally that the resonance frequency will always be $\nu_0$,
by showing that the
interaction hamiltonian is invariant under
rotations of the Bloch vector. 
The interstate s-wave
interaction at point ${\bf r}$ is described by the second-quantized 
hamiltonian density
%%% introduced: at point r
\begin{equation}
H_{\rm int}({\bf r}) = \V\,\psi^\dagger_1({\bf r})\, \psi^\dagger_2({\bf r})\,
\psi^{ }_2({\bf r})\,  \psi^{ }_1({\bf r}) . \label{eq:Hint}
\end{equation}
Under a general rotation, described by polar angles $\theta,
\phi$, the field operators $\psi^\dagger_{1,2}$ transform according to:
\begin{eqnarray}
\psi^\dagger_{1 \theta, \phi} &= &\cos{\mbox{$\frac{\theta}{2}$}} \; {\rm e}^{-{\rm i} \phi / 2} \;\psi^\dagger_1
+ \sin{\mbox{$\frac{\theta}{2}$}} \;{\rm e}^{{\rm i} \phi / 2}  \;\psi^\dagger_2\nonumber\\
\psi^{\dagger}_{2 \theta, \phi} &= -
&\sin{\mbox{$\frac{\theta}{2}$}} \;{\rm e}^{-{\rm i} \phi / 2}\;
\psi^\dagger_1 + \cos{\mbox{$\frac{\theta}{2}$}} \;{\rm e}^{{\rm i} \phi / 2} \; \psi^\dagger_2
\label{eq:rotfield}
\end{eqnarray}
Using the standard fermionic anticommutation relations ($\psi_1
\psi_2 = -\psi_2 \psi_1, \psi_1 \psi_1 = 0$ etc.~), it is
easy to show that:
\begin{eqnarray}
H^{\theta, \phi}_{\rm int} = \V\psi^\dagger_{1 \theta, \phi} \psi^\dagger_{2 \theta, \phi} \psi^{ }_{2 \theta, \phi} \psi^{ }_{1 \theta, \phi} \stackrel{!}{=} H_{\rm int}
%\nonumber\\
%=&&\V\psi^\dagger_1  \psi^\dagger_2  \psi^{ }_2
% \psi^{ }_1  = H_{\rm int}
\end{eqnarray}
We therefore see that an RF-induced rotation on the Bloch sphere commutes with the interaction hamiltonian,
and hence does not change the energy of the many-body state.
It is then obvious that the resonant frequency will always
be $\nu_0$, independent of the coherence of the system.

We now present a more general calculation of the
mean-field frequency shifts, which holds for both fermions and
bosons. 
To reduce complexity and concentrate on the only
controversial case of interstate interactions, we consider a
fictitious boson with no intrastate interactions
($a_{11}=a_{22}=0$). The (local) mean-field expectation value of the
hamiltonian density in Eq.~\ref{eq:Hint} is~\cite{fett71}
%%% introduced: local ... density
\begin{eqnarray}
\label{eq:Eint}
{\cal E}_{\rm int}({\bf r}) = \langle H_{\rm int} \rangle &=& \V (n_{1} n_{2} + \epsilon \;n_{12} n_{21}),\nonumber\\
g^{(2)} &=& 1 + \epsilon \frac{n_{12} n_{21}}{n_1 n_2},
\end{eqnarray}
where $n_1=\langle \psi_1^\dagger\, \psi^{ }_1 \rangle$ and $n_2=\langle \psi_2^\dagger\, \psi^{ }_2 \rangle$ are
the local
%%% introduced: local
densities in the two states, we have introduced ``coherences"
$n_{12}=\langle \psi^\dagger_1\, \psi^{ }_2 \rangle$ and $n_{21}=\langle
\psi^\dagger_2\, \psi^{ }_1 \rangle$, and $\epsilon=\pm 1$ for
bosons/fermions. 
In a fully coherent sample $n_{12}n_{21} = n_1
n_2$ and $\g = 1+\epsilon$. As decoherence sets in, $\g$ increases (decreases) from 0 (2)
to 1 for fermions (bosons).
%In both
%cases, we recover the expected result of Eq.~\ref{eq:meanfield},
%$\langle H_{\rm int}\rangle={\cal E}_{\rm int}$.
For the most general case of a partially decohered sample, 
we can rewrite Eq.~\ref{eq:Eint} in terms of the (local) Bloch vector, using
%%% introduced: local
$n_{1,2} = \frac{n}{2} \mp m_z$, $n_{12} = m_x + i\, m_y = n^*_{21}$, and
$n_{12} n_{21} = m^2_x + m^2_y = m^2_\perp$, where $n$ is the total particle density. This gives
\begin{eqnarray}
{\cal E}_{\rm int} &=& \V \frac{n^2}{4} + \epsilon \V \left|\vect{m}\right|^2 - (1 + \epsilon) \,\V\,m^2_z.\label{eq:EintBloch}
\end{eqnarray}

%%% Changed: Two states with same numbers

Two samples with same numbers of atoms in
%%% inserted: states
states $\one$ and $\two$, but different levels of coherence, have the same $m_z$, but different $|\vect{m_{\perp}}|$ (e.g. states D and E in Fig.~\ref{fig:bloch}b). Again we see that two such samples indeed have different interaction energies.

%%% changed: two such states 

Now, let us evaluate the effect of coherence on the resonant RF frequency.   
A coherent RF excitation preserves entropy ($|\vect{m}|={\rm const.}$), and the total density $n$. In an infinitesimal tilt of the Bloch vector, the density of atoms transferred from $\one$ to $\two$ is $d n_2 = - d n_1 = d m_z$. Therefore, the change of interaction energy per transferred particle, and thus the shift in the resonant frequency $\Delta \nu$, comes out to be
\begin{eqnarray}
\Delta \nu =  \frac{1}{h}\frac{\partial {\cal E}_{\rm int}}{\partial m_z}\bigg|_{n,\,|\vect{m}|} = \frac{1}{h}
(1+\epsilon)\V\,\; (n_1 - n_2) \label{freqshift}.
\end{eqnarray}

In analogy with a spinning top which precesses in the gravitational field, the resonant frequency for an infinitesimal tilt of the Bloch vector is also equal to the frequency of its free precession. In the traditional language of atomic physics, this analogy just reiterates that
Rabi~\cite{gupt03} and Ramsey~\cite{harb02} spectroscopy measure the same characteristic
frequency of the system.
The striking result is that in contrast to the interaction energy (Eqs.~\ref{eq:Eint},~\ref{eq:EintBloch}), the precession of the Bloch vector,
or equivalently the RF frequency shift (Eq.~\ref{freqshift}), {\it does not} depend on the level of coherence in the
sample. Remarkably, the final state may have a value of $\g$ different from the initial state, such that the energy difference per transferred particle is {\it independent} of the initial $\g$.
Equation~\ref{freqshift} explains both our measurements with fermions, and the experiment with thermal bosons of ref.~\cite{harb02}.
%%% changed "dependent" to depend
\begin{figure}[htbp]
  \includegraphics[width=0.8\columnwidth]{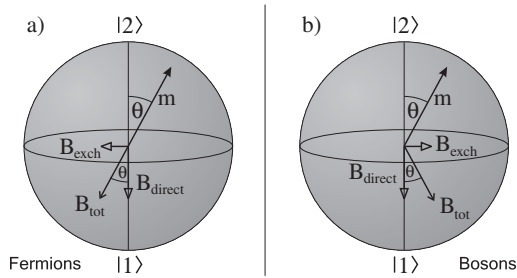}
 \caption{Mean-field represented as effective magnetic field. a) Fermions: The
exchange and direct interaction add up to form a magnetic field
aligned with the average spin ($\V<0$ in the drawing). The net torque vanishes and the
Bloch vector $\vect{m}$ precesses at the unperturbed frequency
$\nu_0$. b) Bosons: The exchange interaction has opposite sign to that in
fermions. It exerts a torque on the average spin equal to the torque
induced by the direct interaction,
as can be seen by comparing
the two cross products with $\vect{m}$. The Bloch vector thus
precesses at $\nu_0$ plus twice the frequency shift due to direct
interaction.}

%%% changed: b) Bosons: The exchange interaction has opposite sign than for fermions.

  \label{fig:meanfield}
\end{figure}

In order to further elucidate the role of coherences in the precession of the Bloch vector, 
we employ the interpretation of the mean-field energy as the interaction of the average spin with an effective magnetic field~\cite{fuch02,fuch03}. 
Using Eq.~\ref{eq:EintBloch}, we obtain ${\cal E}_{\rm int} = {\rm const.} - \frac{1}{2}\vect{B}_{\rm eff}\cdot\vect{m}$~\cite{selfint} with 
\begin{eqnarray}
\vect{B}_{\rm eff} = 2 \V\, (m_z \hat{\vect{e}}_z - \epsilon\, \vect{m}_\perp) \label{eq:Bint}.
\end{eqnarray}
In this picture, the precession of the spin due to interactions is driven by the torque $\vect{B}_{\rm eff} \times \vect{m}$.
The magnetic field along the $z$ axis is induced by the direct interaction, and has the same sign for fermions and bosons (Fig.~\ref{fig:meanfield}). 
The transverse magnetic field comes from the exchange interaction, and has opposite signs for fermions and bosons.
For fermions, $\vect{B}_{\rm eff}$ is parallel to $\vect{m}$ (Eq.~\ref{eq:Bint}) and hence does not cause any precession. Equivalently, the direct and exchange interaction exert torques equal and opposite to each other.
For bosons, the two contributions add up to yield exactly twice the precession frequency given by the direct interaction alone. During decoherence, the exerted torque shrinks in proportion to the decaying transverse spin. Therefore, the precession frequency remains constant, no matter how small the coherences are.

%%% deleted: In addition to the effects discussed here, interactions in inhomogeneous samples can lead to spin
%%% waves~\cite{lewa02, mcgu02, okte02spin, fuch02, will02} both in bosonic and fermionic gases.

%--- end : conclusions -------------------------------------------------

In conclusion, we have demonstrated the absence of the mean-field
shift of RF transitions in a fully decohered,
interacting binary mixture of fermions. This was explained by proving the invariance of the interaction energy under
coherent Hilbert space rotations. This result
is relevant for the potential use of a fermionic atom supplying the frequency standard in an atomic or optical
clock, since it implies a robust elimination of the systematic
errors due to density dependent frequency shifts.
Previously, the absence of such clock shifts was explained by the absence of mean-field energy in a purely coherent superposition state~\cite{gupt03}. Now we have shown that there is no spectroscopic shift even after decoherence has led to measurable mean-field energies. Further, we have
presented a simple theoretical framework for calculating the
precession frequency of the Bloch vector which describes an arbitrary
spin state of either fermions or bosons. This resolves ``The Mystery of the Ramsey Fringe that
Didn't Chirp"~\cite{harb02ICAP} with a simple and intuitive picture.

We thank Claudiu Stan and Christian Schunck for experimental assistance, and Michele Saba for critical reading of the manuscript.
This work was supported by the NSF, ONR, ARO, and NASA.

\end{document}